# Emergence and criticality in spatiotemporal synchronization: the complementarity model

Alessandro Scirè

University of Pavia
Department of Electrical, Computer and Biomedical Engineering
via A. Ferrata 5 – 27100 – Pavia (Italy)
Email: alessandro.scire@unipv.it

**Abstract**

This work concerns the long-term dynamics of a spatiotemporal many-body deterministic model that exhibits emergence and self-organization, and which has been recently proposed as a new paradigm for Artificial Life. Collective structures emerge in the form of dynamic networks, created by bursts of spatiotemporal activity (avalanches) at the edge of a synchronization phase transition. The spatiotemporal dynamics is portraited by a motion picture and quantified by time varying collective parameters, which revealed that the dynamic networks undergo a "life cycle", made of self-creation, self-regulation, and self-destruction. The power spectra of the collective parameters show 1/f power-law tails, and the statistical properties of the avalanches, evaluated in terms of their size and durations, show power laws with characteristic exponents in agreement with those values found in the literature concerning neural networks. The mechanism underlying avalanches is discussed in terms of local-to-collective excitability. Finally, the connections that link the present work to self-organized criticality, neural networks and artificial life are discussed.

## 1. Introduction

In his Metaphysics, Aristotle wrote: "*In the case of all things which have several parts and in which the totality is not, as it were, a mere heap, but the whole is something beside the parts...*" [1], creating the first historical reference to the concept of *emergence*, which saw its more significant developments much later, in the 19th century, by the so-called British Emergentists. One of those, the philosopher G. H. Lewes, coined the term "emergent" in 1875 distinguishing it from the mere "resultant" [2], to describe a compound system that shows behaviors that its parts do not have on their own, and that appear only when such parts interplay in a wider whole. The iconic example was (and still is) the phenomenon of Life, which can be regarded as an emergent property of chemistry, because it is not a property of any component of that system, but is still featured by the system as a whole. British Emergentism reached its apogee thanks to C. D. Broad's *The Mind and Its Place in Nature* in 1925, but the popularity of the emergentist vision waned shortly after Broad's writings due to important scientific developments that smoothened the boundaries between adjacent disciplines, such as quantum chemistry and molecular biochemistry, encouraging more structural explanations for the biological processes [3]. Similarly, in the 1950s the discovery of the DNA as well as the first prebiotic experiments tended to push the concept of emergence away from the phenomenon of Life. In parallel, logical positivists such as C.G. Hempel and T. Nagel turned emergence into a pure epistemic concept [4]. However, the attention to emergence saw a turn in the 1970s, with the disclosure of complex systems of both natural and artificial kind. A large amount of literature accounting for emergence appeared since, motivated by a large variety of systems of different nature. More specifically, emergence was rediscovered and acquired a central role in complex systems physics within the framework of order/disorder phase transitions [5].

In physics, the term order (disorder) traditionally designates the presence (absence) of some kind of symmetry and/or correlation in a many-body system, typically ordered at low temperatures and, after heating, undergoing one or several phase transitions into less ordered states. After assessing the terminology thanks to Paul Ehrenfest's work on statistical mechanics in the early twentieth century [6], the study of phase transitions permeated several physical phenomena ranging from magnetization [7], to liquid crystals [8], Bose–Einstein condensates [9] and superconductivity [10]. More recently, after the availability of computers, scientists were able to experiment with dynamics abstracted from any particular material substrate. This allowed to recast the scheme of phase transition to describe changes between different states of organization in abstract systems, in a way applicable to various disciplines other than physics, including biology [11] and social sciences [12]. Usually, a phase transition is characterized by an order parameter, a collective magnitude that measures the degree of order as a function of an intensive control parameter (e.g., temperature). Order parameters normally range between zero in one phase (usually above the critical point) and nonzero in the other, as sketched in Fig. 1. Close to the critical point is where emergent complex structures are found, at the edge between order and disorder [13]. A process frequently invoked for such emergence, due to the multitude of examples that were reported, is Self-Organized Criticality (SOC).

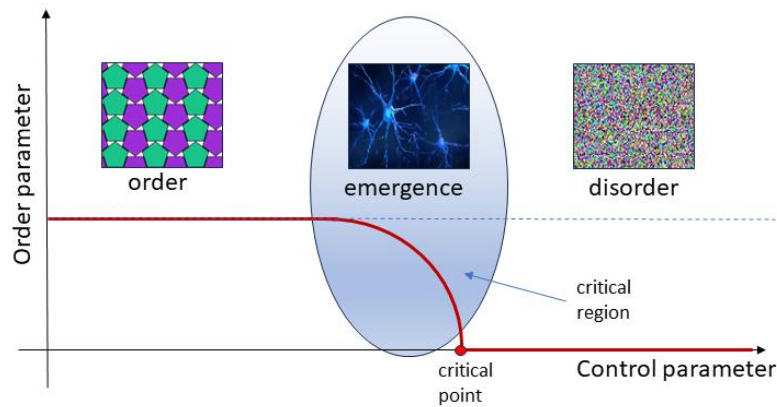

**Fig. 1.** Schematic representation of a generic order-to-disorder continuous phase transition. The red line marks the behavior of the order parameter vs the control parameter. The critical region, where emergence occurs, is emphasized.

The concept of SOC was put forward in the late 1980s [14] and is considered to be one of the general mechanisms by which complexity emerges in nature. It deals with many-unit systems that spontaneously organize their dynamics close to a critical point of a phase transition, by means of scale-invariant abrupt collective events often taking the form of avalanches. SOC found application in fields as diverse as geophysics [15], physical cosmology [16], evolutionary biology and ecology [17], economics [18] and more, including neurosciences. Indeed, the *critical brain hypothesis* [19][20] states that certain biological neural networks function near a phase transitions, and experimental recordings from large groups of neurons contained traces of SOC [21], i.e. bursts of activity, so-called neuronal avalanches, with *1/f* frequency spectra and sizes and durations that follow power law distributions [22]. These results, replicated on a number of settings, supported the hypothesis of a self-organized collective dynamics of a large ensemble of interconnected neurons in the brain close to the critical point of a phase transition, and the phase transition often invoked in the neural networks context is the *collective synchronization phase transition* [23][24], also related to memory [25] and cognition [26].

Besides phase transitions related to thermodynamics, the ubiquitous process of *synchronization* has been connected to phase transitions too. In the late 1960s A. Winfree proposed a coupled oscillators model for the circadian rhythms of the daily activity in plants and animals [27]. He found out that above a critical coupling strength synchronization appears spontaneously, with modalities reminiscent of a phase transition. Later, Y. Kuramoto simplified Winfree's idea in a solvable model [28]. Kuramoto results created a theoretical connection between collective synchronization and phase transitions [29], because in Kuramoto theory an order parameter is defined, and it behaves as in Fig.1 when the diversity of the oscillators (the control parameter) is increased. Kuramoto theory found applications in many interdisciplinary fields, and specifically in large systems of biological oscillators, such as chorusing frogs [30], pacemaker cells [31], and, as mentioned above, firing neurons [32].

Recently, a novel theoretical model for spatiotemporal synchronization, inspired by the Kuramoto model, has been introduced [33]. It consists of a dual ensemble of phase oscillators (possessing *polarity*, a characteristic similar to electric charge), free to move in a two-dimensional space with local interactions, and that in certain conditions [34] show the emergence of complex superstructures. Those superstructures are self-organized dynamic networks, resulting from a synchronization process of many units over length scales much greater than the interaction length. Such networks compartmentalize the two-dimensional space with no a priori constraints, by means of porous transport walls created by critical events, i.e. bursts of activity accompanied by global signals. Those networks emerge in the critical region of a deterministic order-to-disorder phase transition that display static pattern formation, dynamic networks, and irregular dynamics, respectively, for increasing values of the control parameter. However, the asymptotic behavior in the dynamic network regime was left for further investigations, as well as a deeper understanding of the collective excitability properties, and a statistical analysis of those activity bursts.

In the present work, the research of [34] is continued, focusing on the dynamic networks regime and investigating the long-term behavior. The present research unveiled that those networks undergo a "life cycle", made of self-creation, self-regulation, and self-destruction, portraited by motion pictures and quantified by time varying collective parameters. The power spectra of the collective parameters are calculated, showing *1/f* power-laws. The statistical properties of the activity bursts (avalanches) are evaluated in terms of their size and durations, again showing power laws. The mechanism behind the creation of those transport networks, in terms of a transition from local to collective excitability, is addressed and qualitatively unveiled. Finally, the connections of the outcomes with Self-Organized Criticality, Neural Networks and Artificial Life are discussed.

The Manuscript is organized as follows, section 2 is devoted to the description of the model, section 3 shows the results of a long simulation in the dynamic networks regime. The results are organized in four subsections, namely *Spatiotemporal Dynamics, Collective Parameters and Occupation Maps, Avalanche Statistics,* and *From Local to Collective Excitability.* Finally, Section 4 is devoted to summarize the manuscript and to discuss the results in an interdisciplinary view.

## 2. The Complementarity Model

As introduced in [34], the model is a non-linear dissipative dynamical system that governs the dynamics of N locally coupled limit-cycle oscillators, free to move in a 2D space, hence inhabiting a compound phase space given by the direct product $S^1 \otimes \mathbb{R}^2$ for each oscillator. They are abstract mathematical objects consisting of two complementary kinds of oscillators (poles) arbitrarily labelled as *circles* for negative poles and *squares* for positive poles. The circle and square poles obey a complementary interaction form that makes them tend to minimize the distance in space and maximize the distance in phase when different poles interact (e.g. one circle and one square), or maximize the distance in space and minimize the distance in phase when equal poles interact (e.g. two squares) [33][34]. This led to the choice of *complementarity model* as a name for the following set of differential equations.

The equations of motions, for *N* poles, read

$$\dot{x}_i = \sum_{j=1}^{N} \nabla_i \ W(|x_i - x_j|) \cos(\varphi_i - \varphi_j), \qquad (1)$$

$$\dot{\varphi}_i = \gamma_i \Delta + \sum_{j=1}^{N} \gamma_i \gamma_j \ W(|x_i - x_j|) \sin(\varphi_i - \varphi_j), \qquad (2)$$

where $\nabla_i$ means differentiation respect to $x_i$. The potential wells (that make the interactions *local* with a characteristic length L = 1) are chosen as exponential wells

$$W(|x_i - x_j|) = -e^{-|x_i - x_j|^2}, \qquad (3)$$

where $x_{i,j}$ are the spatial coordinates (in $\mathbb{R}^2$) and $\varphi_{i,j}$ the local phases of the *i,j-th* oscillator with *i,j = 1,... N*, and $|x_i - x_j|$ is the Euclidean distance in $\mathbb{R}^2$. The coefficients $\gamma_i$ define the *i-th* oscillator polarity, i.e. $\gamma_i = 1$ if the oscillator is a square (*s-pole*) or $\gamma_i = -1$ if a circle (*c-pole*). Mathematically, the model (1)-(2) is a dissipative non-linear dynamical system, including one control parameter $\Delta$. The effect of $\Delta$ in Eq. (2) is to split the natural frequencies of *c* and *s*-poles, a symmetry breaking term able to produce dynamical states. See [33] and [34] for more details about this model.

## 3. Results

### *Spatiotemporal dynamics*

As a continuation of the preliminary analysis concerning the regime labelled as "dynamic networks" obtained in a range of values for $\Delta$, longer simulations have been performed in that regime, i.e. close to the critical point of the order/disorder phase transition introduced in [34]. The chosen ensemble for the present numerical integration of Eqs (1)-(2) contains N =1000 poles, 500 circles and 500 squares, and a value of the detuning ($\Delta = 0.1$) that bias the system where the dynamic networks appear. The starting conditions are reported in Fig. 2, and consist of random positions in a $\sim 30 \times 30$ box in $\mathbb{R}^2$ and random phases, uniformly chosen in $[0, 2\pi]$. The phases are as usual [33][34] encoded by the marker colors. The colormap (see Fig. 2) smoothly changes through various colors and it is circular, i.e. the values 0 and $2\pi$ for the phases correspond to the same color. The spatiotemporal dynamics, resulting from the numerical integration of Eqs. (1) - (2) during $3.5 \times 10^7$ integration steps, is reported in the movie <u>S1</u>.

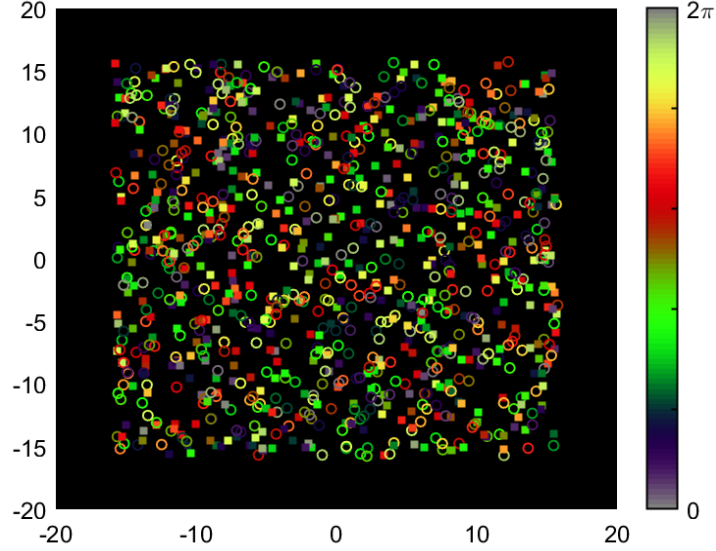

**Fig.2.** Starting Conditions. Uniform random positions for N =1000 poles, 500 circles and 500 squares, in a $\sim 30 \times 30$ box and uniform random phases in $[0, 2\pi]$. The phase color coding is a circular colormap where 0 and $2\pi$ have the same color.

The movie <u>S1</u> reports on a spatiotemporal dynamics where the population splits in two: a "solid" phase, i.e. a matrix of "neutralized" dipoles made of overlapping circles and squares (the solid phase) that progressively acquires a regular conformation (*dipolar lattice*), bounding a co-present dynamic transport "liquid" network. In in the first part of the movie <u>S1</u> the liquid network continuously reshape itself, exchanging units with the solid lattice in a complex dynamics, as shown. Such intricate scenario has been partially disclosed in [34], but the longer simulation time here unveiled that, on the long run, the liquid transport network gradually relaxes onto a topology that changes less dramatically with time, and sustains a quasi-stationary coherent flux and shape, as shown in the central part of movie <u>S1</u>. However, such condition is not definitive, after a long interval where the liquid transport network and the dipolar lattice interplay in a dynamic equilibrium rarely interrupted by small critical events, the liquid transport process loses coherence and breaks locally, as if it were not able anymore to coordinate the dynamics in that region. Immediately afterwards the whole liquid transport loses coherence and disarticulates itself in a set of residual disconnected patterns, as shown in the last part of movie <u>S1</u>.

*Collective Parameters and Occupation Maps*

Consistently with [34], a number collective parameters are considered to achieve quantitative insights in the scenario showed by the movie <u>S1</u>, and to better identify the 3 parts of the dynamics above mentioned. A modified Kuramoto order parameter

$$\rho e^{i\theta} = \frac{1}{N} \sum_{k=1}^{N} \gamma_k e^{i\varphi_k} \qquad (4)$$

allows to quantify the degree of collective synchronization vs time. In particular, $\rho$ measures the *global degree of entrainment,* it takes values between 0 and 1, where 1 means total coherence (global phase locking) and 0 means total disorder (disordered phases, unlocking). The phase $\vartheta$ is the global phase*,* and the *global frequency* $\Omega(t) = d\vartheta/dt$ describes collective pulsations with regular ($\Omega \sim$ constant) or chaotic dynamics ($\Omega$ fluctuating), or collective excitability, when temporal spikes appear in the time traces of $\Omega$ [35]. The total kinetic energy is

$$T = \sum_{k=1}^{N} v_k{}^2, \qquad\qquad (5)$$

where $v_k = \dot{x}_k$ are the spatial velocities, and it quantify the global activity in space. The time traces of T, $\rho$, and $\Omega$ respectively, for the simulation reported in movie <u>S1</u>, are shown in Fig. 3. By inspection, Fig. 3 can be divided in 3 regions, corresponding to different time intervals corresponding to the 3 regimes identified in the movie S1. For each of those regimes the relative occupation maps, i.e. how many times a pole was present in a point in space during that interval, are reported in Fig. 4.

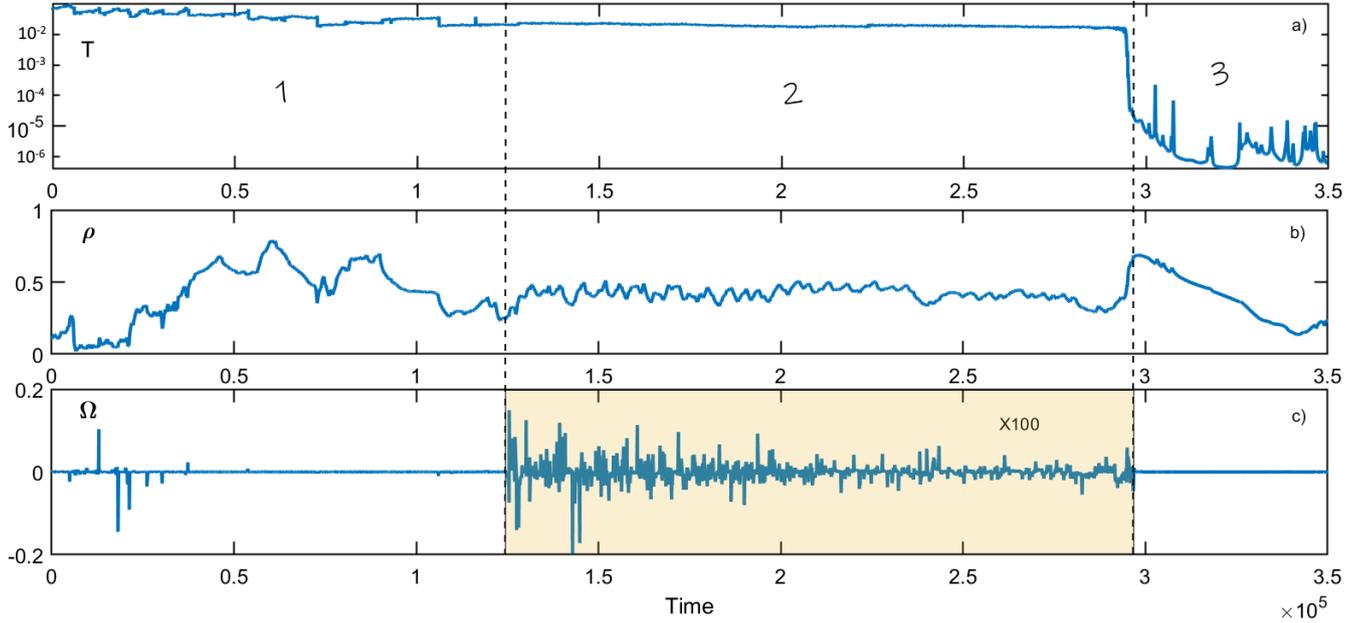

**Fig. 3**. Time traces of the collective parameters: a) total kinetic energy T, b) entrainment $\rho$ and c) the global frequency $\Omega$. Each time step corresponds to 100 integration steps. $\Delta$ = 0.1, N=1000, 500 circles and 500 squares.

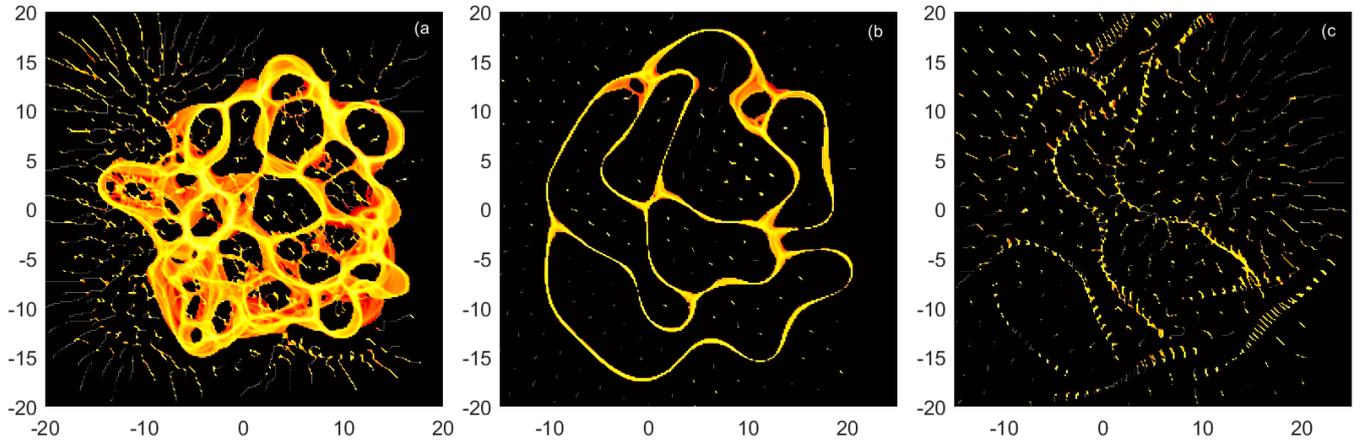

**Fig. 4**. Occupation Maps concerning region 1, 2 and 3 of Fig. 3 (panel a, b and c respectively). a) Network Construction. b) Network Self-Regulation: a two phases regime with compresence of a liquid transport network and a solid lattice. c) Network death. $\Delta$ = 0.1, N=1000, 500 circles and 500 squares.

Follows the analysis of the 3 regimes:

1) *Time*: $0 \sim 1.25{\times}10^5$ – **Region 1** in Fig. 3. <u>Network construction</u>: Is basically the dynamics reported in [34], that shows the formation of a looped dynamic transport network, supporting a counterpropagating flux of $c$ versus $s$-poles, continuously reshaping and reorganizing itself, in an intricate spacetime dynamics, non-trivially coupled to an equally intricate phase dynamics. In this region global phase waves, signalized by flashes of colors that involve many oscillators, are not uncommon. At the same time a regular matrix of neutral dipoles is forming (the solid dipolar lattice), bounding the transport network, and continuously exchanging poles with it. The total kinetic energy T(t) shows irregular jumps and the occupation map (Fig. 4a) shows the reshaping of transport loops of different sizes.

2) *Time*: $1.25 \times 10^5 \sim 3 \times 10^5$ – **Region 2** of Fig.3. <u>Network self-regulation</u>: The total kinetic energy T(t) shows small chaotic fluctuations (see Fig. 5a) interrupted by rare small jumps, drawing a scroll attractor in the $(T, \dot{T})$ plane (Fig. 5b). The entrainment ρ(t) displays small fluctuations close to the center of its range of excursion, balancing between order and disorder. The global frequency Ω(t) shows small irregular bursts. The occupation map (Fig. 4b) concerning this region shows the persistence of a smooth network structure made of connected loops. The transport network here settles to a regime structure where critical events (avalanches) are less intense respect to region 1.

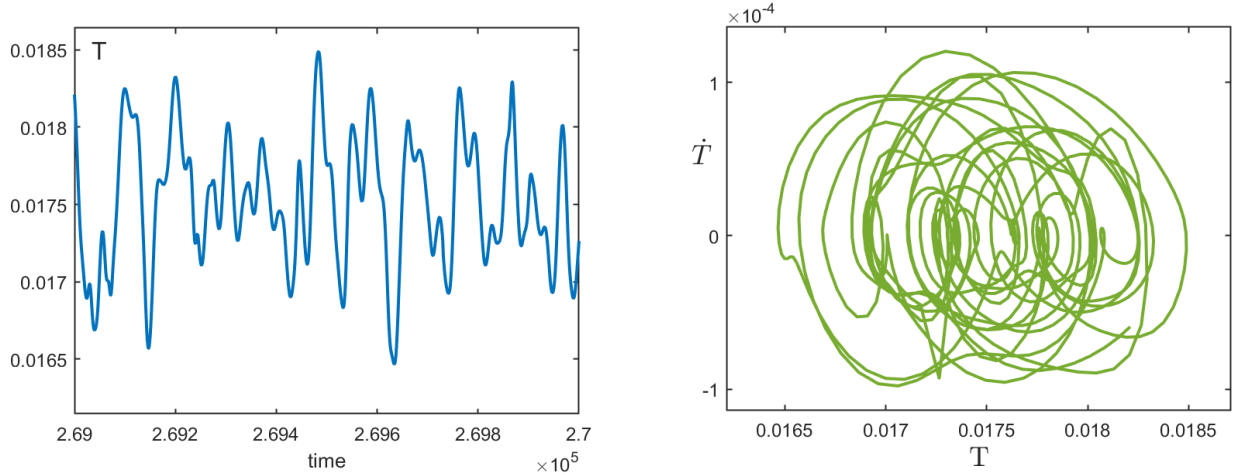

**Fig.5**. <u>Left panel</u>: Time traces of the total kinetic energy T in a small portion of region 2 of Fig.3a (network self-regulation), showing small chaotic fluctuations. <u>Right panel</u>: The scroll chaotic attractor pertinent to the time interval of the left panel.

Both solid and liquid phases are important for the construction of the network, as new connections are formed when a certain number of solid dipoles liquefy creating currents and new connections or, conversely, existing connections are removed when liquid moving poles solidify into the dipolar lattice. During the network self-regulation regime, the solid and liquid subpopulations are well separated in space and time, as shown (Fig. 6) by the histogram of the squared velocities of the ensemble in that regime, reporting two well separated peaks for the solid and liquid phases respectively.

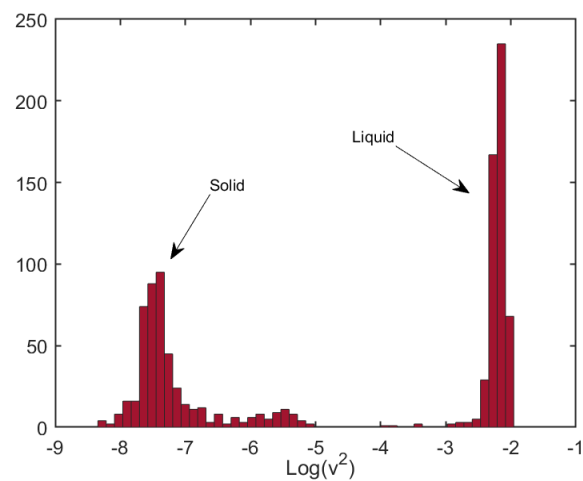

**Fig. 6.** Squared velocities histograms concerning region 2 of Fig. 3 (Network self-regulation) – The solid phase (small velocities) and the liquid phase (higher velocities) are well separated.

3) After $3 \times 10^5$ – <u>Network death</u>: **Region 3** of Fig.3. The total kinetic energy T(t) drops suddenly to small values, interrupted by peaks of residual activity, the entrainment ρ(t) after a quick rise falls towards desynchronization, and the global frequency Ω(t) stops bursting. The associated occupation map (Fig. 4c) shows that the transport network has disappeared leaving behind a few disconnected patterns.

Network death happens by local failures in the flux, that create destructive avalanches that disarticulate the network, leaving sparse patterns showing some residual activity. However, the system can be "defibrillated", i.e. by applying a strong perturbation to the local frequencies, a new transport network can be started.

In region 2 of Fig. 3 the transient of region 1 is extended while creating a smoothly reshaping network, weakly chaotic, that organizes the collective dynamics half the way between coherence and incoherence. When exploring the long-term dynamics, those networks revealed their transitory nature, undergoing a spontaneous destruction process. This scenario is due to a *critical slowing down*, the slow regression to equilibrium typical of processes at criticality, an indicator of neighboring a phase transition well known in many different contexts [36].

The analysis continues calculating the power spectra of the time traces of T(t) and ρ(t) of Fig.3, reported in Fig. 7. The T-spectrum shows a *1/f* low frequency power law tail and a high frequency oscillatory bump, whereas the ρ-spectrum fits the *1/f* "pink noise" power law, over 5 decades. Those kinds of spectra typically appear at criticality in phase transitions, where the large fluctuations in the value of the global parameters are characterized by power-law scaling.

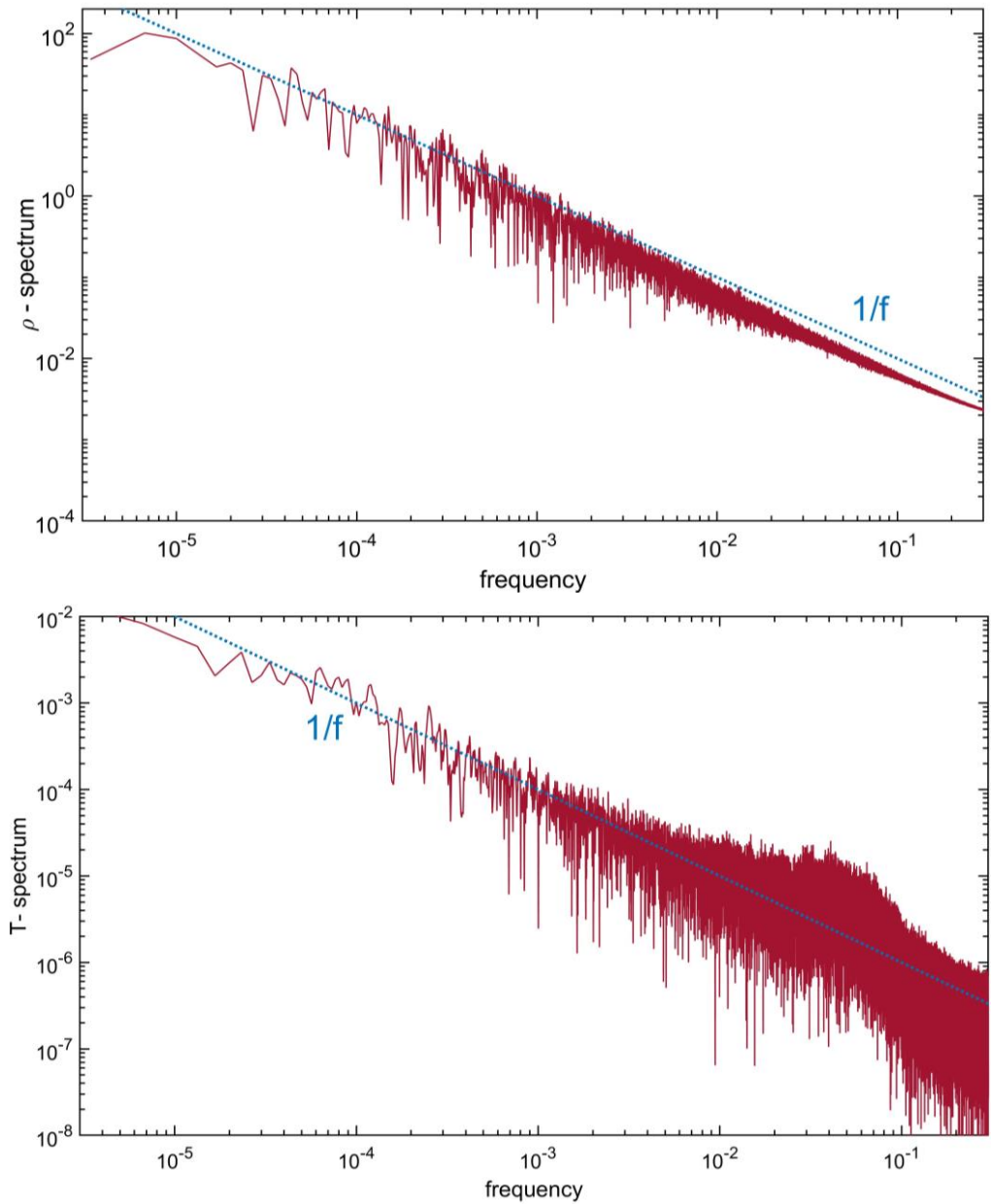

**Fig. 7.** Frequency spectra of the kinetic energy T (upper panel) and of the entrainment ρ (lower panel) concerning the time traces reported in Fig.3. The power law 1/f is reported for reference (blue dots).

*Avalanches statistics*

A closer look to the time traces of $\Omega$ showed an irregular sequence of peaks in the global frequency, a mark of collective excitability [35]. The amplitude of those peaks decreases with time, heavily in region 1 and smoothly in region 2 of Fig. 3. In Fig. 8 (left panel) a sample of the time trace of $\Omega(t)$ in region 2 is reported, showing the details of those peaks of irregular sizes and durations. In [34] those peaks were related to changes in the network structure, avalanches caused by the creation/destruction of transport loops or connections, i.e. periods of intense activity. The bigger the connection created or destroyed the bigger the avalanche, because it involves the coordinated action of many units with a consequent stronger contribute to the order parameter. When computing the frequency spectrum of $\Omega(t)$ in region 2 (see Fig. 8 – right panel) a low frequency power law tail again appears, together with a frequency bump at in the middle-high part of the spectrum (see arrow in Fig. 8).

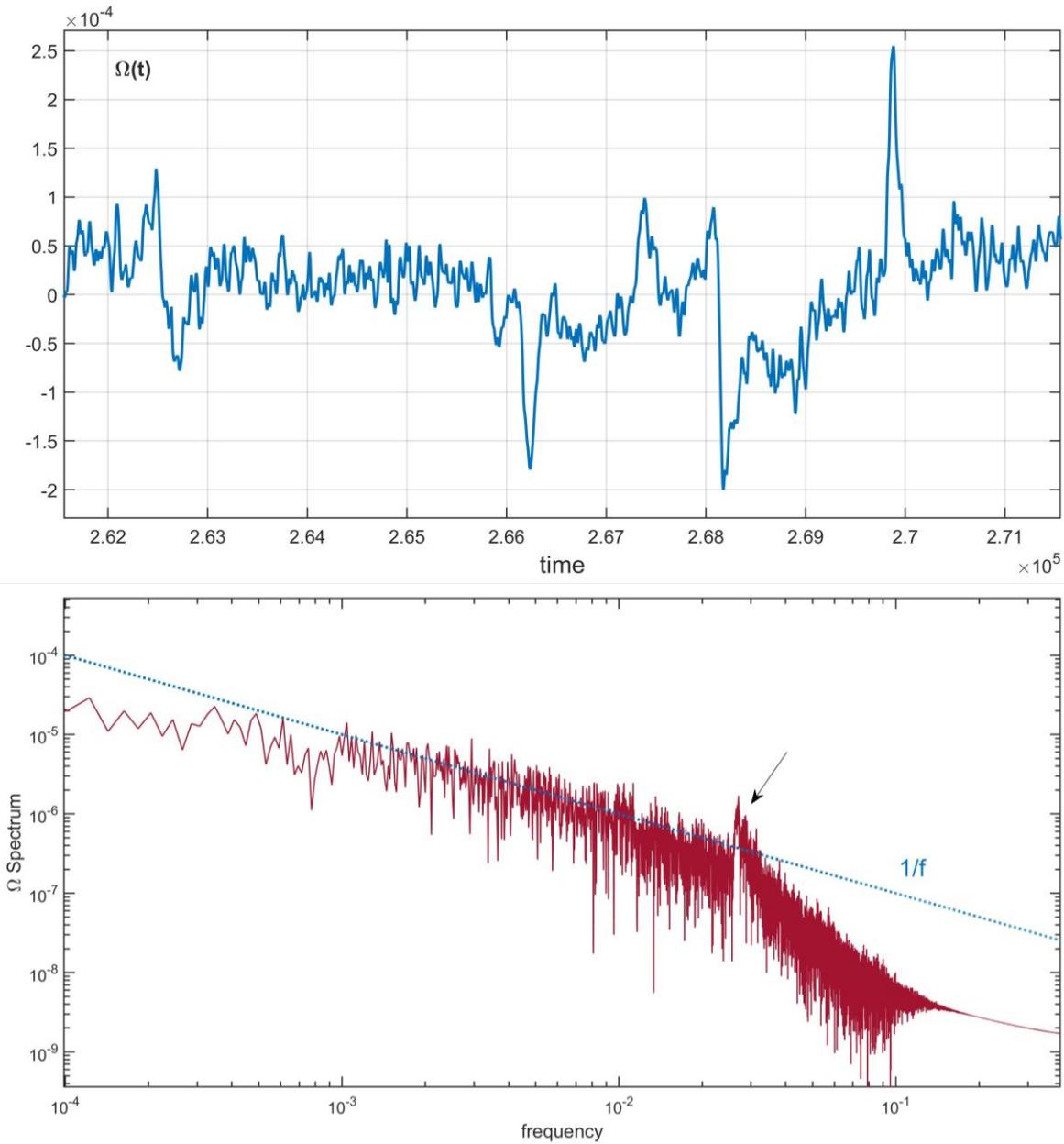

**Fig. 8**. Upper panel: Time trace of $\Omega(t)$ in a small portion of region 2 of Fig. 3. Lower panel: Frequency spectrum of $\Omega(t)$ concerning whole region 2 of Fig. 3.

Moreover, the statistical analysis of temporal $\Omega$-peaks has been performed, in terms of peaks heights and time durations, the results are shown in Fig. 9. The statistics shows power-law distribution of avalanches in both height and size, with power laws exponents in agreement with those value commonly found in the neural networks literature [19-24].

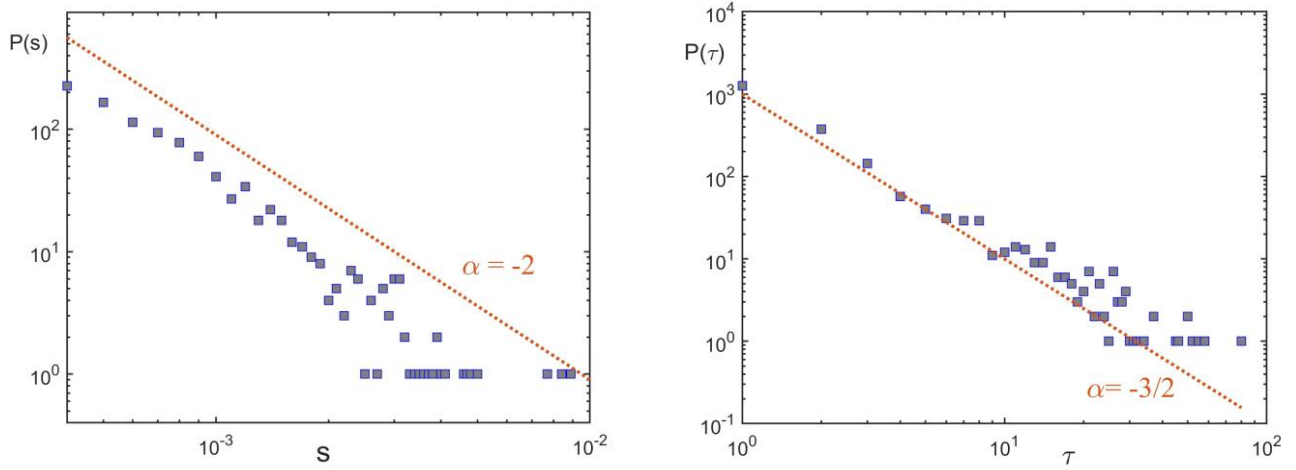

**Fig. 9**. Avalanches statistics concerning the time peaks in the global frequency $\Omega(t)$. Panel a) Avalanches sizes S histogram and power-law $s^\alpha$ with $\alpha = -2$. Panel b) Avalanches durations $\tau$ histogram and power-law $\tau^\alpha$ with $\alpha = -3/2$.

*From local to collective excitability.*

This subsection is devoted to illustrate the excitability properties of the system, at both a local and a collective level. The local excitable unit of system (1)-(2) is represented by the cs-dipole, a dual unit. Indeed, specifying Eqs. (1)-(2) for one s- and one c-pole the problem reduces to a two-dimensional system which admits a potential. The cs-dipole allows for a two-dimensional potential of the form

$$V(x, \varphi) = W(x)\cos(\varphi) - \Delta\varphi, \qquad (6)$$

Where $W(x)$ is the local interaction function (3), $x$ is the relative position of the c-pole to the s-pole, and $\varphi$ is the relative phase. The potential function $V(x, \varphi)$ is reported in Fig. 6a for $\Delta = 0.2$ as an example. The potential can be understood as the Adler potential [37] in $\varphi$, partly modulated by $W(x)$. It contains 2 fixed points, one stable and one unstable, that come closer when increasing $\Delta$, until they coalesce and disappear via saddle-node bifurcation. For $\Delta < 1$ the system is excitable, i.e. it responds with a big excursion in the phase space if the stimulus exceed a certain threshold. In particular for a sufficiently strong spike in $\Delta$ a large oscillation in space takes place (see Fig. 6, panels b and c), separating the dipole for a certain amount of time.

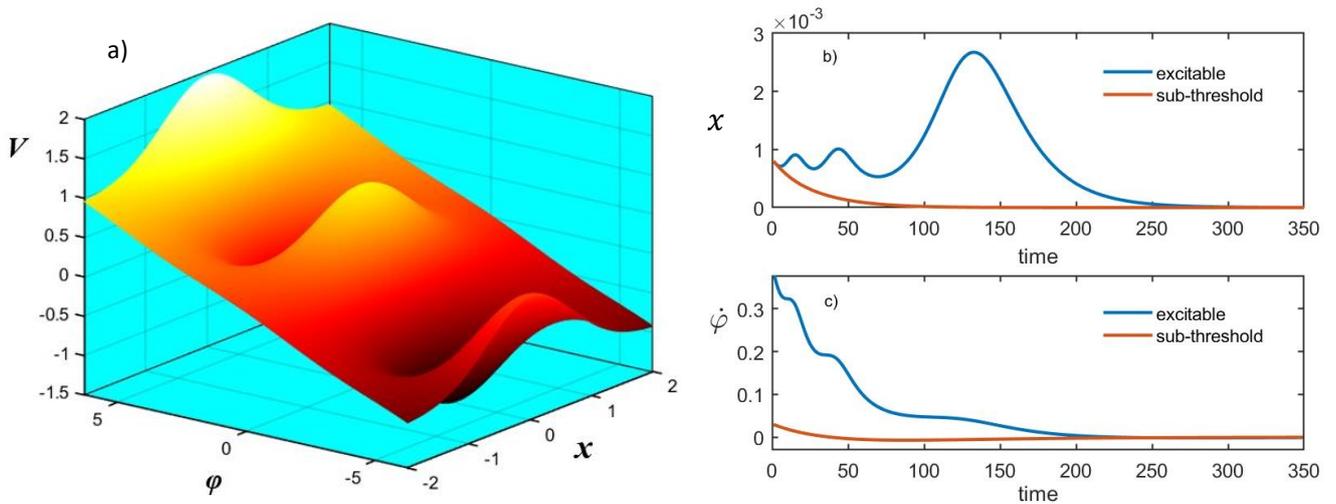

**Fig. 10.** Panel a) The two-dimensional dipolar potential of Eq. (6) plotted for $\Delta = 0.2$. Panels b) and c) Response to a spike in $\Delta$ stimulus. If the stimulus overcomes a threshold the system shows a large excursion in $x$.

The step from local excitability to avalanches that create the counterpropagating currents shown in the movie S1 can be understood as follows. Considering a set of dipoles, e.g. set in a row as in Fig. 11. A perturbation able to split one dipole will propagate along the chain, pushing poles of the same kind and pulling poles of the opposite kind. Such push/pull dynamics propagates the perturbation along the chain creating counterpropagating currents. So, when for some reason a cs-dipole is excited locally, the local excitation is transmitted to the neighboring elements which are excited in turn, creating avalanches without preferred space scales. Such avalanches that can follow closed pathways in loops of any size, creating spatial self-similarity and scale free statistical and spectral properties.

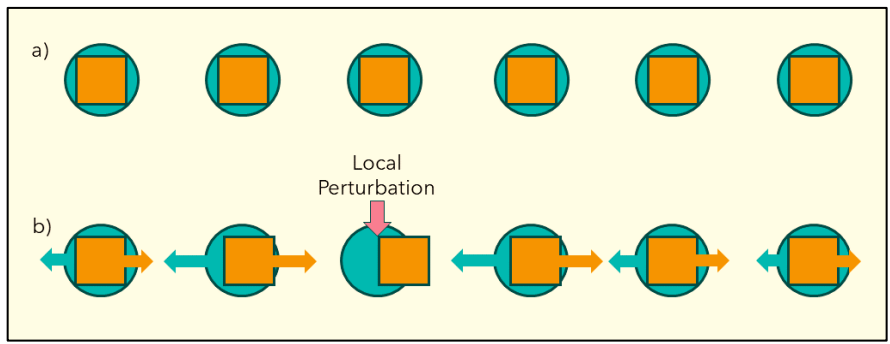

**Fig.11**. Schematic representation of how a local perturbation can propagate along a chain of dipoles.

## 4. Discussion

This work concerns the long-term emergence and self-organization properties of a recently introduced deterministic, dissipative many-body system for spatiotemporal synchronization. The system describes the dynamics of many dual agents (poles), phase oscillators free to move in a plane, each given a positive or negative polarity as a further degree of freedom. The poles interact according to a complementarity interaction scheme, hence the name *Complementarity Model*. In a previous work, the system was shown to undergo an order-to-disorder phase transition when a control parameter was varied. As is common in phase transitions at criticality, order (solid) and disorder (liquid) phases coexist, but in this case such compresence has peculiar properties from the dynamical, statistical and organizational point of view. Close to the critical point, the system creates a liquid transport network that percolates over the whole system size and displaying a sort of "life cycle", made by the sequence of a self-creation, self-regulation, and self-destruction stages. Self-creation is characterized by a turbulent dynamics, rich of intense critical events, that create both a liquid network and a solid lattice, competing for space dominance by means of excitable avalanches that continuously create or damp the transport currents that feed the network. Such currents define connections and loops of different sizes and velocities in the network, and, if damped, feed in turn the solid lattice. The collective degree of synchronization shows strong and irregular variations, and so does the kinetic energy. However, as time increases, the intensity of the critical events decreases until a self-regulation stage follows: a highly organized collective spatiotemporal behavior, where the liquid network and the solid lattice *complement* each other in a self- sustained global shape that stems from a dynamic equilibrium, interrupted by sparse critical events of weak intensity. This continues until such equilibrium breaks abruptly, and the laboriously achieved structure self-destructs revealing its transient character, typical of those constructs at the edge of a phase transition. One more product of the critical slowing down, but with the capacity to extend its existence by organizing the dynamics half the way between order and chaos in an autonomous way.

The results contained in the present work can be interpreted as a form of self-organized criticality for the following reasons: 1) The process of creation of the dynamic networks relies on critical events (avalanches) that promotes local excitability to the collective scales. 2) Avalanches have no preferred spatial size, forming connected loops with a wide distribution of sizes. 3) The frequency spectra of the collective parameters show power laws, notably close to the *1/f* law commonly observed in biological systems. Moreover, the exponents of the power laws concerning the sizes and durations of the avalanches are in agreement with the corresponding values largely found in the neural networks literature concerning neural spiking activity. These facts, together with having the emergence of a complex plastic network at the edge of a synchronization phase transition, shaped by avalanches that propagate excitable waves over long distances, inevitably suggests a connection with neural networks and with the critical brain hypothesis. However, the present work is not aimed at modelling the brain or any material entity, it aims at modelling a *process*, not a physical structure. Such process concerns *how* complex, coherent, self-regulating and unique collective structures emerge from the local interaction of simple and undifferentiated oscillating bodies. Besides, *uniqueness* is an important feature displayed by the present model. Indeed, a small change in the starting conditions would create a different network with a different life cycle. This happens due to the well-known properties of chaotic systems, despite the constitutive dipoles are all the same. So, *diversity*, a central issue in biology, here spontaneously emerges as a collective feature, without invoking mutations or random changes.

The phase transition in this *deterministic* complementarity model is not driven by disorder, but by a control parameter that acts more like a global bifurcation parameter rather than as a "temperature". When analyzed in both a statistical and a dynamical sense, the complementarity model exhibits key features shared by both phase transitions and global bifurcations to chaos. A similar scenario is displayed by deterministic cellular automata [38]. Indeed, the "life cycle" of the liquid network represents a new and different example of what C. Langton (one of the founders of the field of Artificial Life [39]) defined as *extended transient dynamics* in cellular automata, and that he advocated to be the artificial equivalents of the precursors of the biological activity, the prebiotic forms [40]. "*One of the most exciting implications of this point of view is that life had its origin in just these kinds of extended transient dynamics. Looking at a living cell, one finds phase-transition phenomena everywhere. The point of view advocated here would suggest that we ourselves are examples of the kind of "computation" that can emerge in the vicinity of a phase transition given enough time. Now nature is not so beneficent as to maintain conditions at or near a phase transition forever. Therefore, in order to survive, the early extended transient systems that were the precursors of life as we now know it had to gain control over their own dynamical state. They had to learn to maintain themselves on these extended transients in the face of fluctuating environmental parameters, and to steer a delicate course between too much order and too much chaos […]*". So how could this structures evolve? Two or more spatially separated superstructures could create a community by mutual synchronization, because each superstructure as a whole can be seen as a chaotic oscillator, again susceptible of synchronization (the same process that created it) in a bigger scale. The same process can hence repeat itself, climbing space scales while creating a cascade of nested evolutive layers of increasing complexity, steering the evolution between order and chaos. Evolution would therefore rely on processes that are coordinated in both time and space: the extremely rapid transient flows (very short pulses of chemicals) are propagated to longer and longer time scales of minutes, hours, days, and so on via interlocking processes. Processes that would permit attaining highly complex form of interactions and social ties such as the human ones, because those interactions, however complex, are still traceable in a form of synchronization, as shown by neural synchrony in social and affective neuroscience [41].

* * *

**Acknowledgements**

Looking forward to have that barricaded grappa together, A.S. wishes to acknowledge Prof. Valerio Annovazzi-Lodi (University of Pavia) for supportive conversations.

**Supporting Information Captions**

S1 - MOVIE – Spatiotemporal dynamics obtained from the numerical integrations of Eqs (1)-(2) during $3.5 \times 10^7$ integration steps with N =1000 poles, 500 circles and 500 squares, and $\Delta = 0.1$. The starting conditions are reported in Fig. 2, and consist of random positions in a $\sim 30 \times 30$ box in $\mathbb{R}^2$ and random phases, uniformly chosen in $[0, 2\pi]$. The phases are encoded by the marker colors. The colormap is reported in Fig. 2.

**Declarations**

**Ethical Approval**
Not Applicable.
**Competing Interests**
The author has declared that no competing interests exist.
**Author's Contribution**
A.S. performed all the simulations, calculations, and he wrote and reviewed the Manuscript.
**Funding**
The author received no specific funding for this work.
**Availability of data and materials**
All relevant data are within the paper and its supporting information files.